# Evaluation of Mg compounds as coating materials in Mg batteries

Tina Chen,[1,2] Gerbrand Ceder,[1,2] Gopalakrishnan Sai Gautam,[3,#] and Pieremanuele Canepa[4,*]

[1]Department of Materials Science and Engineering, University of California Berkeley, California 94720, United States.
[2]Materials Science Division, Lawrence Berkeley National Laboratory, Berkeley, California 94720, United States.
[3]Department of Mechanical and Aerospace Engineering, Princeton University, Princeton, New Jersey 08544, United States.
[4]Department of Materials Science and Engineering, National University of Singapore, Singapore 117575, Singapore.

[#]Corresponding author: gautam91@princeton.edu
*Corresponding author: pcanepa@nus.edu.sg

## Abstract:

Mg batteries utilizing a Mg metal anode with a high-voltage intercalation cathode define a potential pathway toward energy storage with high energy density. However, the realization of Mg batteries is plagued by the instability of existing electrolytes against the Mg-metal anode and high-voltage cathode materials. One viable solution to this problem is the identification of protective coating materials that could effectively separate the distinct chemistries of the metal-anode and the cathode materials from the electrolyte. Using first-principles calculations we map the electrochemical stability windows for non-redox-active Mg binary and ternary compounds in order to identify potential coating materials for Mg batteries. Our results identify Mg-halides and $Mg(BH_4)_2$ as promising anode coating materials based on their significant reductive stability. On the cathode side, we single out $MgF_2$, $Mg(PO_3)_2$ and $MgP_4O_{11}$ as effective passivating agents.

## 1. Introduction

Multivalent batteries, such as those based on Mg, present a potential alternative to Li-ion batteries, particularly in terms of increased energy density.[1] Mg batteries are able to use Mg metal as an anode at reasonable current densities (< 0.5 $mA/cm^2$),[2] which in combination with the higher oxidation state of Mg (+2 rather than Li's +1) can provide a significant increase in the energy density of Mg batteries compared to Li-ion batteries. So far, prototypes of Mg batteries have utilized electrolytes, such as $MgCl_2$ with $AlCl_3$, $Mg(ClO_4)_2$, $Mg(NO_3)_2$, $Mg(TFSI)_2$ and more complex molecules dissolved in acetonitrile, THF, or glymes-based solvents, in combination with Mg metal as the anode and a low voltage sulfide cathode ($Mg_xMo_6S_8$ and $Mg_xTiS_2$).[2-15]

Typical Mg electrolytes have significantly narrower electrochemical stability windows (~1.5 V-3.0 V vs. Mg)[16] compared to what is available in the Li-ion battery space (~1.5 V-5 V vs. Li).[17] Indeed, most electrolytes, including the solvents used in commercial Li electrolytes (e.g., PC and DMC),[18] have poor reductive stability (i.e., cathodic stability) and tend to decompose at the Mg metal anode.[6, 19] In addition, the utilization of high-voltage cathodes (e.g., oxides) is greatly impeded by the limited oxidative stability (i.e., anodic stability) of Mg electrolytes.[2-7, 9-14, 20-22] Thus, the reactivity of the electrolyte against both Mg-anode and a high-voltage cathode results in electrolyte decomposition, often producing a passivating layer primarily containing a binary Mg-salt, such as MgO (and $Mg(OH)_2$ if moisture is present).[23-26] The presence of MgO greatly inhibits $Mg^{2+}$ transport[27] and eventually the ability of the battery to store energy reversibly.[28] Further work is still being done to develop Mg electrolytes that can reversibly strip and deposit Mg at the anode and cathode.[5, 6, 29] For example, a class of carboranes has recently been proposed as promising electrolytes stable against Mg metal and high voltage cathodes (up to 4.6 V vs. Mg).[7] However, more work is required to elucidate the mechanisms of reversible Mg transfer at the cathode and develop strategies to mitigate electrolyte decomposition.[30, 31]

In analogous Li-systems, several approaches have been utilized to address the safety and electrochemical stability limitations of typical Li electrolytes.[32, 33] For example, solid electrolytes have been shown to be safer compared to typical solvent-based electrolytes, which may experience thermal runaway issues.[34-37] Another ongoing field of research is the application of protective coating layers to shield one or both electrodes from an incompatible electrolyte, while providing sufficient ionic mobility and preferably low electronic conductivity. Indeed, the solid electrolyte interphase (SEI) that forms at the graphitic anode-electrolyte interface is a good example of a protective layer with sufficient Li mobility that enables the reversible operation of Li-ion batteries.[38] Therefore, similar solutions can be envisioned for Mg-batteries as well. To accomplish this goal, we search for materials that can act as either protective coatings or even solid electrolytes by analyzing the electrochemical stability of various Mg-containing compounds.

Using a combination of density functional theory (DFT) calculations and thermodynamics, we assess the electrochemical stability of various Mg-binary and ternary compounds, which may form as a result of electrolyte decomposition at either the Mg-metal anode or a high-voltage cathode. Specifically, we consider all Mg binaries and ternaries that do not contain redox-active metal ions (except $Ti^{4+}$) and that are known to be electronic insulators. The choice of Mg compounds is also motivated by the highly reducing conditions that appear when in contact with Mg metal. For example, Li binaries and ternaries, such as $Li_3N$, $Li_3P$, LiH, $Li_2S$, $Li_2O$, and LiCl tend to form (and be stable) at the Li electrolyte-anode interface in Li-ion batteries.[39]

By calculating the electrochemical stability windows of candidate compounds, we identify their oxidative and reductive voltages. Our findings provide general guidelines for developing, via either *in situ* or *ex situ* deposition techniques, protective coating materials that are compatible with the anode or the cathode or both. Provided good bulk $Mg^{2+}$ mobility exists,[40] some of these materials may be investigated as protective coating materials or even solid electrolytes.

## 2. Methodology

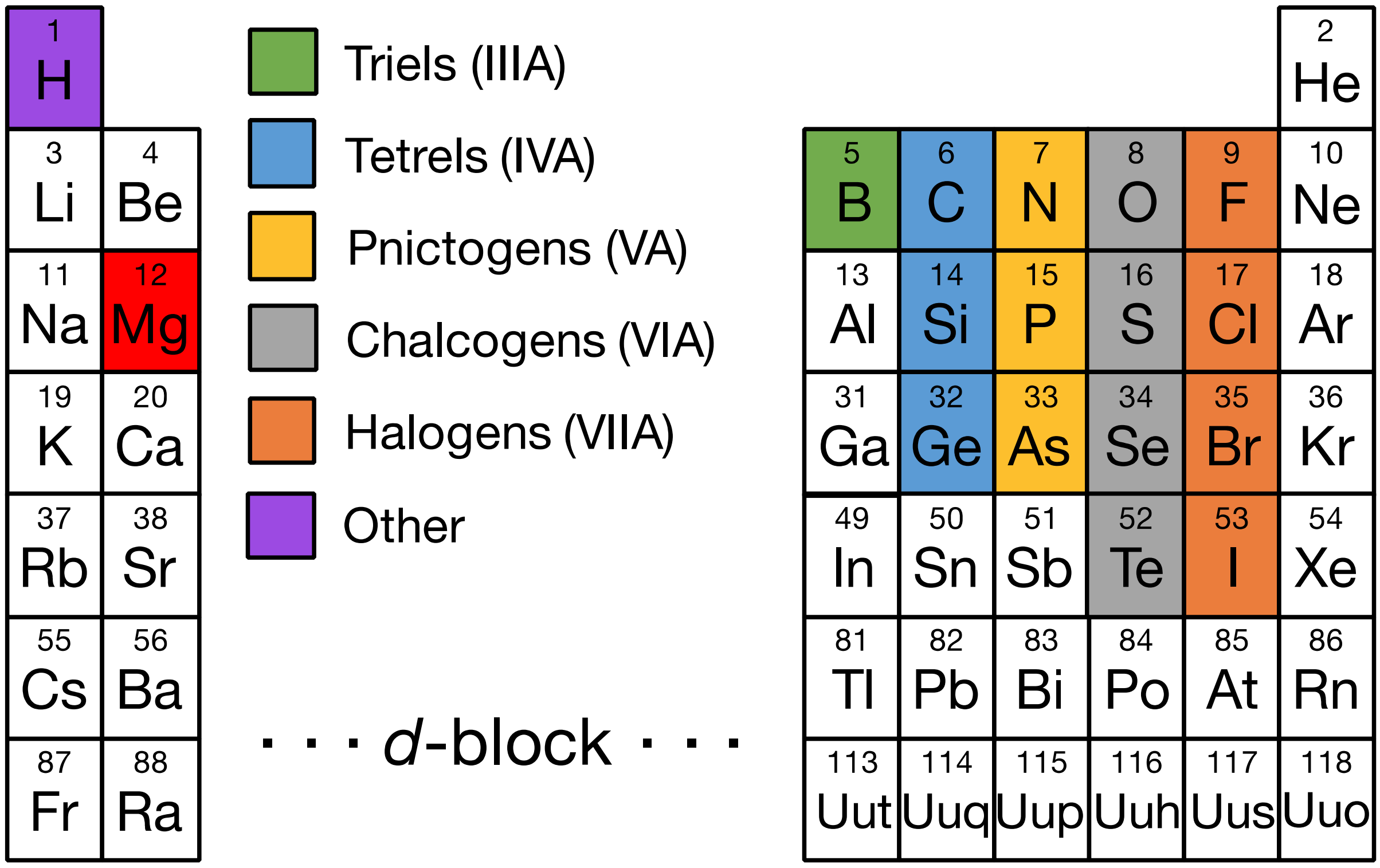

**Figure 1**: Periodic table highlighting the non-transition-metal elements that form binary (and ternary) compounds with Mg (red), including triels (Group IIIA, green), tetrels (Group IVA, light blue), pnictogens (Group VA, yellow), chalcogens (Group VIA, gray), halogens (Group VIIA, orange) and other elements (Hydrogen, purple). We considered all Mg-X binaries and stable Mg-X-Y ternaries, where X and Y are highlighted elements, with the exception of the Mg-X-H chemical space where only Mg-B-H compounds were considered. In addition, we evaluated some compounds containing a non-Mg metal, such as Sc, Ti, Nb, Zr, Al, Ga, and In, either because they are commonly used as coating materials in Li-ion batteries or have been considered as Mg ionic conductors in prior studies.

The set of elements from which we evaluate Mg binaries and ternaries is shown in Figure 1, with Mg colored in red and the other elements colored based on their respective group numbers (a complete list of all Mg-binaries and ternaries investigated is provided in Table S1 of the Supporting Information – SI). In addition to the highlighted elements, we considered borohydrides, niobates, titanates, titanium phosphates, and zirconium phosphates which have been reported to be promising coating materials in Li-ion batteries.[39] Also, we included Mg-(Sc/In)-(S/Se) compounds since they have been explored as potential Mg solid-electrolyte materials in prior studies,[27, 41] apart from Mg-(Al/Ga/In)-(O/S/Se).

The electrochemical stability windows of each compound are calculated using the approach developed by Richard *et al.*[39] by constructing the corresponding grand potential ($\phi$) phase diagram by means of the pymatgen library,[42, 43] where $\phi$ is defined as:

$$\phi[c, \mu_{Mg}] = E[c] - n_{Mg}[c] * \mu_{Mg} \quad (1)$$

For all $\mu_{Mg}$, we construct the convex hull in the grand potential composition-space and identify compounds that are stable at each $\mu_{Mg}$. The Mg chemical potential $\mu_{Mg}$ relates directly to the voltage vs. $Mg/Mg^{2+}$ via Eq. 2:

$$V = -\frac{\mu_{Mg}}{zF} \quad (2)$$

where $F$ is the Faraday constant, $z$ is the number of electrons transferred ($z$ = 2 for Mg) and $\mu_{Mg}$ is referenced to the energy of Mg metal. The internal energy of each compound ($E$ in Eq. 1), in the relevant chemical space, is either obtained from the Materials Project[42, 44] database or is calculated directly using DFT[45, 46] (see Section S2 in SI for more details on the calculation parameters used). For each compound, we utilize the atomic coordinates reported in the Inorganic Crystal Structure Database (ICSD)[47] as initial guesses during our DFT structure relaxation. For $Mg_{0.5}Zr_2(PO_4)_3$ and $Mg_{0.5}Ti_2(PO_4)_3$, which are disordered structures in the ICSD database, we enumerated possible configurations within the respective unit cell[43, 48-50] and included the lowest energy configuration.

## 3. Results and Discussion

*Electrochemical stability windows of Mg-binaries*

Figure 2 shows the voltage windows of all Mg-X binaries considered, where the compounds are grouped by the anion column (Figure 1) and sorted within each group by increasing electronegativity.

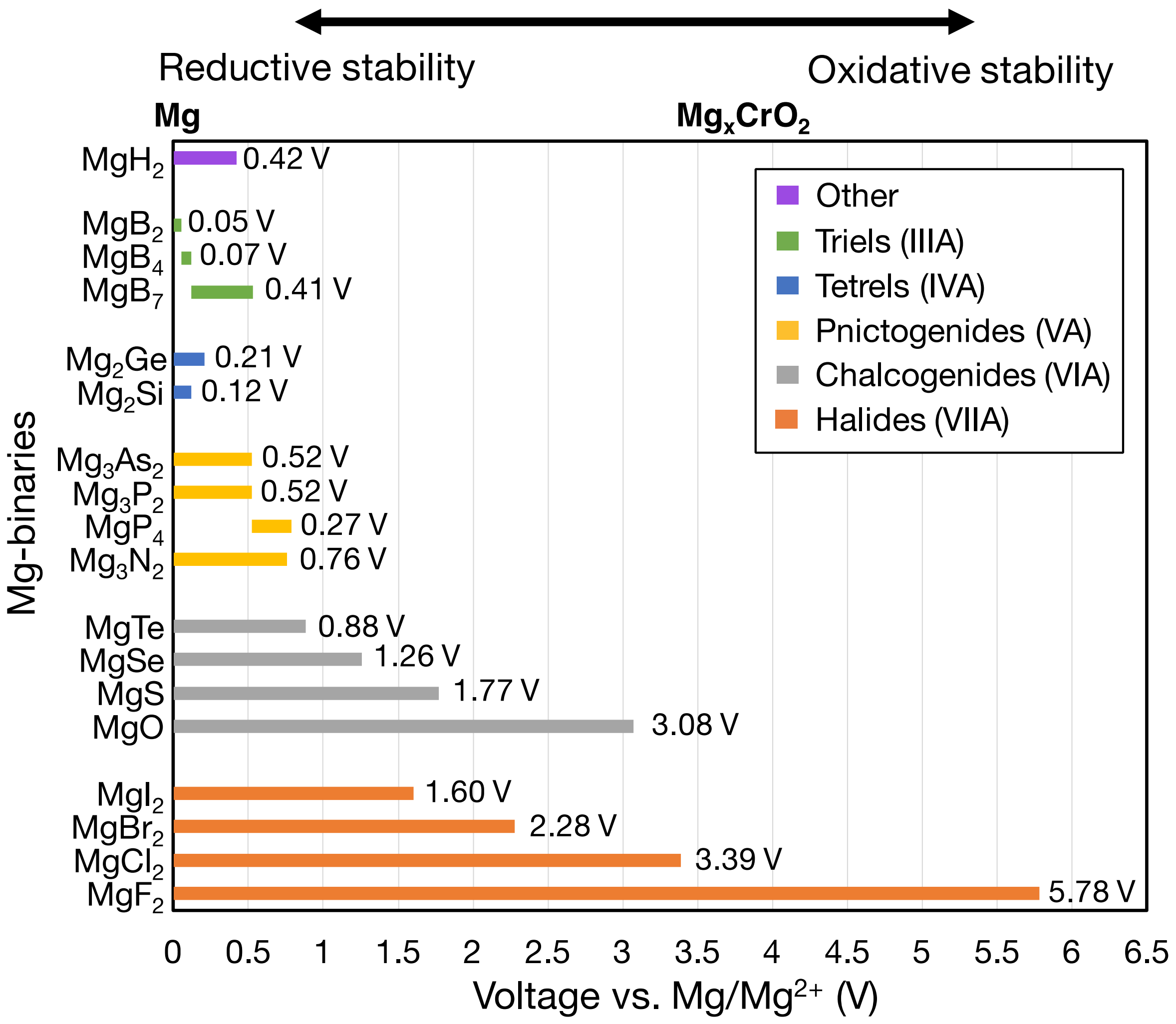

**Figure 2**: Electrochemical stability windows of non-metallic Mg-binaries, indicating the voltages (vs. Mg metal) at which the compound is stable against decomposition. Compounds that are not stable at any voltage, such as the Mg-carbides, are not shown. Compounds are grouped by the anion column and compounds within each group are ordered by increasing electronegativity. For systems with multiple compositions, compounds are ordered by decreasing ratio of Mg to anion. The number at the end of each bar indicates the width of the voltage window. The $Mg_xCrO_2$ spinel is shown above the plot at its calculated average voltage (~3.6 V) for reference.

For a binary system with multiple stable compounds (e.g., Mg-B), we order them according to a decreasing ratio of Mg to anion (Mg:B). Only binaries that are thermodynamically stable (i.e., with negative formation energy at 0 K) are shown. Unstable compounds have been removed from Figure 2 because they will not be stable at any $\mu_{Mg}$. For example, $MgC_2$ has a formation energy of 173 meV/atom at 0 K. The left and right ends of the bar for each compound indicate the lower and upper voltage limits, corresponding to the reductive (cathodic) and oxidative (anodic) stabilities, respectively. Lower reductive stabilities and higher oxidative stabilities imply better resistance against reduction and oxidation, respectively. Thus, the width of the bar (text annotation to the right of each bar in Figure 2) for a given compound signifies its electrochemical stability window. The zero on the voltage axis is referenced to bulk Mg metal (i.e., V vs. $Mg/Mg^{2+}$). Higher voltage values mimic the open circuit voltages of cathode materials, such as Chevrel-$Mo_6S_8$ (~ 1.1 V),[8] layered-$V_2O_5$ (~3.3 V)[51] or $Mg_xCrO_2$ (~3.6 V).[20]

Of note, all of the Mg-halides, Mg-chalcogenides, and Mg-pnictides (except $MgP_4$) are stable at 0 V vs. $Mg/Mg^{2+}$ and thus stable against Mg metal. Among the Mg-triels and Mg-tetrels, only $MgB_2$, $Mg_2Ge$, and $Mg_2Si$ are stable vs. Mg metal. However, the widths of the stability windows of $MgB_2$, $Mg_2Ge$, $Mg_2Si$ are small (< 0.1 V), and thus Mg-triels and Mg-tetrels do not appear to be viable coating materials against typical electrolytes. The poor stability windows of $MgB_2$, and $Mg_2Ge$, $Mg_2Si$ may be attributed to the weak electronegativity of the anions (i.e., B, Ge, and Si) and a consequent low resistance to oxidation. Additionally, B forms three thermodynamically stable compounds at various oxidation states with Mg, namely $MgB_2$ (oxidation state of B is –1), $MgB_4$ ($B^{-0.5}$), and $MgB_7$ ($B^{-0.28}$). While $MgB_2$ is stable against Mg metal (highest reducing conditions), at increasing voltages (~0.05 V vs. $Mg/Mg^{2+}$), compounds with higher B oxidation states become stable, limiting the oxidative stability of $MgB_2$. On the other hand, Cl and Mg only forms $MgCl_2$ as a stable binary, which oxidizes directly to $Cl_2$ gas at ~3.39 V vs. $Mg/Mg^{2+}$. Notably, $MgCl_2$ is used as a precursor for Mg-Al-Cl-based electrolytes and its limited solubility in an ether-based solvent (typically used in Mg batteries) is well documented.[3, 4] Therefore, $MgCl_2$ may already be present in existing electrolytes, given its stability against Mg-metal (Figure 2), and may inherently protect the anode against further reactions with the electrolyte. In light of this, the role of $MgCl_2$ as a potential protective coating layer on the Mg metal electrode needs to be further investigated.

Within each group of compounds of Figure 2 (i.e., each column of Figure 1), there is a strong correlation between the electronegativity of the anion and the oxidative stability. For example, within halogen compounds (orange bars), the oxidative stability rigorously follows the order $MgF_2 > MgCl_2 > MgBr_2 > MgI_2$, which correlates with the relative order of electronegativity of F > Cl > Br > I. Analogous trends can be observed among chalcogens (gray bars) and pnictogens (yellow). From this analysis we conclude that the electronegativity of the anion can be used as a proxy for the oxidative potential of Mg binary compounds since it describes the ability of the anion to limit an oxidation reaction.

*Electrochemical stability windows of Mg-ternaries*

Figure 3 shows the voltage windows of Mg ternary and quaternary oxides, while Figure 4 shows the voltage windows of Mg ternary non-oxides (i.e., sulfides, selenides, tellurides, and a hydride).

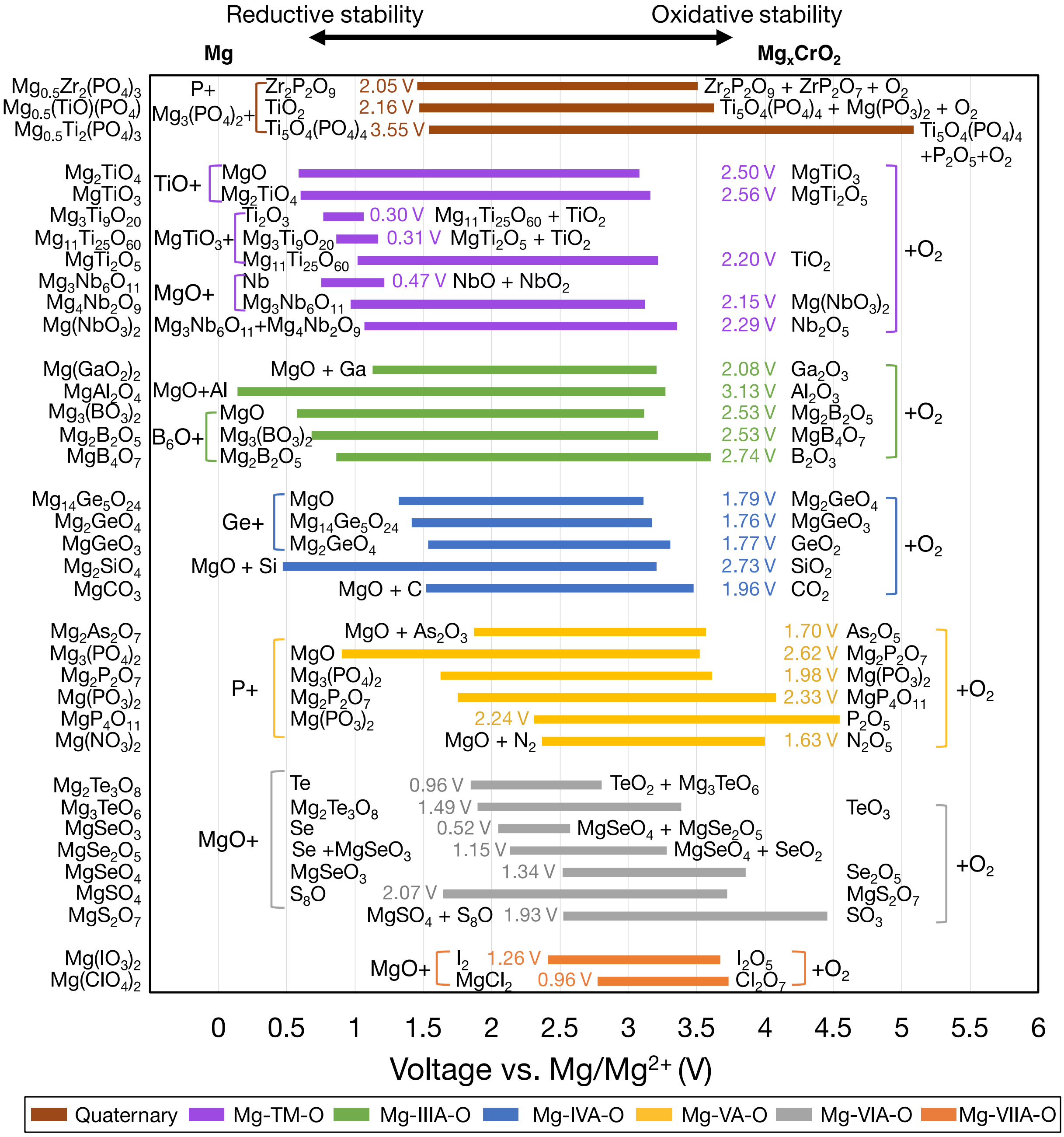

**Figure 3**: Electrochemical stability windows of Mg-ternary and quaternary oxides, indicating the voltages (vs. Mg metal) at which the compound is stable against decomposition. Compounds that are not stable at any voltage are not shown. Ternaries are grouped by the periodic table column of the non-Mg, non-anion elements and ordered within each group by increasing electronegativity of the non-Mg cation. For systems with multiple compositions, compounds are ordered by increasing reductive stability. The text next to each bar indicates the width of the voltage window and the decomposition products at the reductive and oxidative limits. Compounds sharing common decomposition products, such as MgO or $O_2$ are grouped together by brackets.

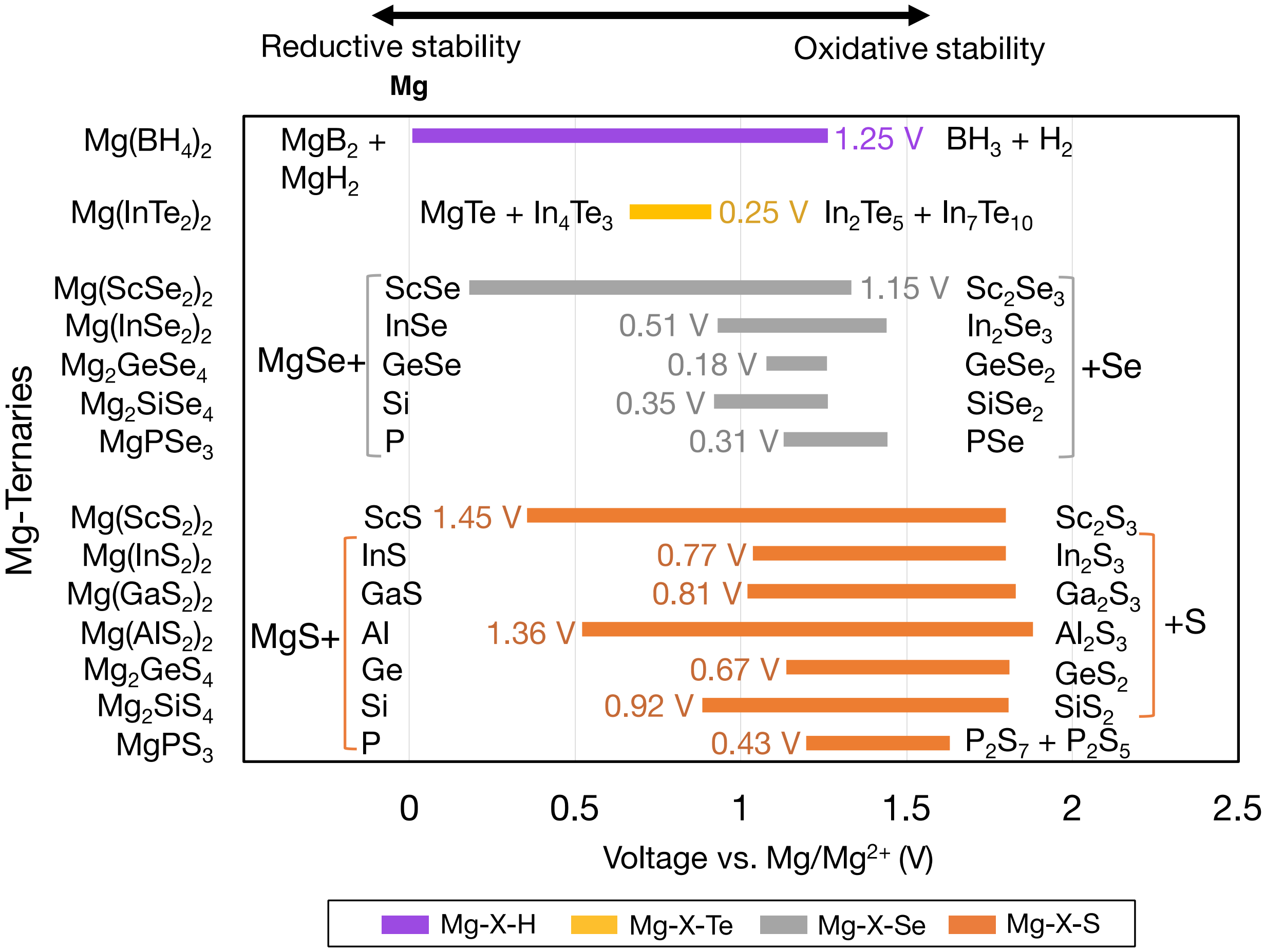

**Figure 4**: Electrochemical stability windows of Mg-ternary non-oxides, indicating the voltages (vs. Mg metal) at which the compound is stable against decomposition. Compounds that are not stable at any voltage are not shown. Ternaries are grouped by anion in order of increasing electronegativity and ordered within each group by increasing electronegativity of the non-Mg cation (e.g., P, S.). The text next to each bar indicates the width of the voltage window and the decomposition products at the reductive and oxidative limits. Compounds sharing common decomposition products such as MgS or S are grouped together by brackets.

The widths of the voltage windows are written next to the respective bars on either the left or right side. Decomposition products at the reductive (oxidative) stability limits are written to the left (right) of the bars. For compounds within a group that share a common decomposition product (such as, MgO, $O_2$ in Figure 3, and MgS, S or MgSe, Se in Figure 4), the common compounds are factored out and indicated by brackets. The compounds shown are based on the elements highlighted in Figure 1 and a set of stable Mg-niobates, Mg-titanates, Mg-titanium-phosphates and Mg-zirconium-phosphates are plotted vs. $Mg/Mg^{2+}$ as the reference. Compounds that are not thermodynamically stable (i.e. with a non-zero decomposition energy or energy above the convex hull) are not plotted. For example, $Mg_{14}Si_5O_{24}$ is calculated to decompose into $Mg_2SiO_4$ and MgO and hence not included in Figure 3. Of note, Mg does not form ternary halides where the halogen is the anion, according to the structures available in the ICSD. Thus, no compounds in the ternary phase spaces of Mg-P-Cl, Mg-N-F, Mg-B-F are known to exist. Instead, we find that the stable Mg-ternaries are ternary chalcogenides, where the anion is oxygen, sulfur, selenium, or tellurium (except for the Mg-borohydride).

Based on Figures 3 and 4, we observe that Mg ternaries do not show reductive stability against Mg metal, as indicated by the lack of reductive stability down to 0 V for any compound considered. The ternary with the best reductive stability is $Mg(BH_4)_2$ (purple bar in Figure 4), which is stable up to 0.01 V against Mg metal. Further, none of the ternary compounds exceed the anodic stability of $MgF_2$ (~5.8 V, Figure 2). Among the ternaries, the Mg-B-O-based compounds, $Mg_2SiO_4$ and $Mg_3(PO_4)_2$ have the widest stability windows, with voltage window widths greater than 2.5 V. Additionally, there exist a few ternary oxides, such as $MgP_4O_{11}$ (~ 4.55 V), $MgS_2O_7$ (~ 4.45 V), and a quaternary $Mg_{0.5}Ti_2(PO_4)_3$ (~ 3.82 V) which have significantly high oxidative stability and may represent potential protective coatings for high-voltage oxide cathodes.[20]

In general, trends in oxidative and reductive stability from Figures 3 and 4 can be explained by analyzing the species most prone to oxidation and reduction, respectively. In most of the ternary compounds considered, the species most prone to oxidation is the anion since the other elements are already at their highest oxidation states (e.g., $P^{5+}$ in $Mg_3(PO_4)_2$). Thus, the susceptibility of the anion to be oxidized dictates the oxidative stability of the compound. For example, among the phosphates (yellow bars in Figure 3), thio-phosphates (orange bar in Figure 4), and seleno-phosphates (gray bar in Figure 4), phosphates exhibit the highest oxidative stabilities compared to $MgPS_3$ and $MgPSe_3$ because $O^{2-}$ is more difficult to oxidize than $S^{2-}$ or $Se^{2-}$. Given that the electronegativity of the anion directly corresponds to the tendency of the anion to attract electrons and its resistance to oxidation, there is a high degree of correlation between increasing anion electronegativity (e.g., O > S > Se)[52] and higher oxidative stabilities of binary (Figure 2) and ternary (Figures 3 and 4) oxides compared to sulfides and selenides, respectively. Additionally, the hybridization of the anion (e.g., between $O^{2-}$ and $P^{5+}$ in $PO_4^{3-}$ moieties) tends to stabilize it by lowering the energy of its electronic states, making the anion more difficult to oxidize. For example, binary MgO, where $O^{2}$ hardly hybridizes with $Mg^{2+}$, oxidizes at ~3.10 V vs. Mg. On the other hand, most Mg-ternary oxides (including the phosphates) oxidize at higher voltages (i.e., exhibit superior oxidative stability) due to the hybridization of the $O^{2-}$ by the non-Mg cation, such as $P^{5+}$, $S^{6+}$, etc.

The reductive stability of ternary compounds depends primarily on two key metrics: *i*) the electronegativity of the species that undergoes reduction, which is the non-Mg cation in ternary compounds, and *ii*) the electronegativity of the anion that does not undergo reduction but regulates the thermodynamic stability of the ternary compound versus the corresponding binary compounds. Notably, reductive stability correlates inversely with the electronegativity of the non-Mg cation species, since larger electronegativities reflect higher attraction towards electrons and a higher propensity for reduction. For example, the reductive stability of ternary compounds (Figure 3) follows the order Mg-Cl-O (~ 2.78 V vs. Mg) < Mg-S-O (~1.65 V) < Mg-P-O (~ 0.9 V) < Mg-Si-O (~ 0.47 V) < Mg-Al-O (~0.14 V), which is the inverse of the electronegativity trends, namely Cl (3.16) > S (2.58) > P (2.19) > Si (1.90) > Al (1.61).[52] In the case of quaternary systems, such as Mg-Ti-P-O and Mg-Zr-P-O, we predict that $P^{5+}$ reduces in preference to $Ti^{4+}$ and $Zr^{4+}$ (brown bars in Figure 3), which is consistent with the larger electronegativity of P (2.19) vs. Ti (1.54) and Zr (1.33).

Importantly, higher electronegativity of the anion results in poorer reductive stability of the ternary compound. For example, the reductive stability among Mg-Ge-, Mg-Sc-, Mg-In-ternary oxides follow Mg-Ge-O (~1.32 V) < Mg-Ge-S (~1.13 V) < Mg-Ge-Se (1.08 V), Mg-Sc-S (~0.36 V) < Mg-Sc-Se (0.18 V), and Mg-In-S (~1.04 V) < Mg-In-Se (~0.93 V) < Mg-In-Te (~0.67 V), respectively, consistent with the anion electronegativity trend (O > S > Se > Te). Note that higher anion electronegativity leads to more stable Mg-binary compounds, i.e., Mg-binaries with larger stability windows (Figure 2), which are common decomposition products under reducing conditions. A more stable Mg-binary reflects a larger thermodynamic driving force for reduction, as quantified by the corresponding formation energy (MgO ~ –3.06 eV/atom, MgS ~ –1.76 eV/atom, MgSe ~ –1.25 eV/atom, and MgTe ~ –0.87 eV/atom),[44] resulting in a lower reductive stability. Interestingly, the compound with the highest reductive stability, $Mg(BH_4)_2$, is composed of a low electronegative anion and a non-Mg cation, H (2.20) and B (2.04), respectively. Thus, minimizing the electronegativities of both the non-Mg-cations and the anions could be the key to discovering ternary compounds that are stable against Mg-metal.

Notable exceptions to the aforementioned trends in reductive stability vs. (non-Mg cation/anion) electronegativity can be observed across different chemistries in Figures 3 and 4. For example, electronegativity of B (2.04) > Ga (1.81) > Al (1.61), but the reductive stability of Mg-Al-O (~0.14 V) > Mg-B-O (~ 0.58 V) > Mg-Ga-O (~ 1.13 V). Similar trends can be observed among Mg-IVA-O, and Mg-VA-O compounds (Figure 3). Such anomalies can be attributed to two factors that override non-Mg-cation electronegativity trends: *i*) stability of Mg-(IIIA/IVA/VA) binaries (signifying the thermodynamic driving force to form decomposition products), and *ii*) the relative position of the empty electronic states of IIIA/IVA/VA elements, as influenced by the extend of hybridization with oxygen (difficulty in reducing the ternary compound). For example, the highest oxidative stability of binary Mg-Al alloys (~0.06 V,[44] not shown in Figure 2) is lower than both Mg-B compounds (~0.53 V, Figure 2) and Mg-Ga alloys (~0.19 V, not shown). On the other hand, the significant hybridization of the electronic states of P with O likely pushes the empty (anti-bonding) P states to higher energy levels, making P difficult to reduce in ternary Mg-P-O, compared to As in Mg-As-O and N in Mg-N-O.

In the case of reductive stability vs. anion electronegativities, the stability of Mg-Al-O (~0.14 V) > Mg-Al-S (~ 0.52 V), and Mg-P-O (~ 0.9 V) > Mg-P-S (~1.20 V), despite the electronegativity of O > S. Here, the discrepancy can be attributed to the stability of Al-O and P-O bonds in comparison to Al-S and P-S bonds, as quantified by the formation energies ($Al_2O_3$ ~ –3.44 eV/atom, $Al_2S_3$ ~ -1.46 eV/atom and $P_2O_5$ ~ –2.46 eV/atom and $P_2S_5$ ~ –0.64 eV/atom).[44] The higher stability of Al-O and P-O bonds is possibly due to better hybridization of Al and P among the oxides versus sulfides, respectively. Thus, despite MgO creating a larger thermodynamic driving force for reduction than MgS (as indicated by the stability windows in Figure 2), the lack of affinity for S from Al and P in Mg-Al-, and Mg-P-ternaries facilitates the reduction of $Al^{3+}$ and $P^{4+/5+}$, respectively, in the ternary sulfides compared to the oxides.

*Potential candidate materials*

Based on the voltage windows of the Mg binaries, ternaries and quaternaries in Figures 2-4, we suggest potential coatings on both the Mg metal//Mg electrolyte and the Mg electrolyte//cathode interfaces. At the cathode interface, the oxidative stability should be high for candidate compounds. Among the binaries, only $MgF_2$ has an oxidation limit above 4.0 V, whereas among the ternaries, including $Mg(PO_3)_2$, $MgP_4O_{11}$, $Mg(NO_3)_2$ and $MgS_2O_7$ show oxidation limits above 4.0 V. Note that among the candidate materials, those with the widest voltage windows should be given preference, which may enable compatibility with liquid electrolytes that are stable against Mg metal. Therefore, among the high-oxidation-limit compounds, $MgF_2$, $Mg(PO_3)_2$, $MgP_4O_{11}$ and $Mg_{0.5}Ti_2(PO_4)_3$, which have the widest voltage windows (all > 2.0 V), should be considered the most promising candidate materials.

For the Mg metal//Mg electrolyte interface, the reductive stability of a candidate compound should ideally be ~0 V vs. Mg metal. In this context, $Mg(BH_4)_2$, with a reductive stability of ~0.01 V vs. Mg is a promising candidate for a protective anode coating. Previous experiments utilizing $Mg(BH_4)_2$-containing electrolytes have reported the formation of a Mg-conducting interphase layer against Mg-metal with an oxidative stability of 1.7 V vs. Mg, which agrees generally with our computational results (1.25 V vs. Mg).[11, 53] The higher oxidative stability of $Mg(BH_4)_2$ observed by experiments than theory could be due to kinetic stabilization, which is not accounted for in our calculations. Thus, $Mg(BH_4)_2$ should be further investigated as a protective coating on the Mg-metal anode. Additionally, in scenarios where the reductive stability is < ~0.5 V, such as $MgAl_2O_4$, $Mg_2SiO_4$ (Figure 3), $Mg(ScS_2)_2$, and $Mg(ScSe_2)_2$ (Figure 4), the compounds may exist in a metastable manner and may still be valid candidates. For example, in Li-ion batteries the solid electrolyte, garnet-$Li_7La_3Zr_2O_{12}$, has an estimated reductive stability of ~0.1 V vs. Li but has been shown to be metastable against Li metal.[39, 54] However, recent theoretical and experimental studies have shown that $Mg(ScS_2)_2$ and $Mg(ScSe_2)_2$ tend to decompose to binary MgS/MgSe and ScS/ScSe against Mg metal, ruling out any metastable existence.[27, 41] Another case to consider is when the Mg metal anode is replaced by Bi (or Sb or their alloys) as the reductive potential of the anode is shifted by up to ~+0.32 V vs. Mg metal.[55] In the case these alternative anodes are used, several coating materials, such as $MgAl_2O_4$ or $Mg(ScSe_2)_2$, could be envisioned as potential coating materials. Nevertheless, changing the anode chemistry can not only change the overall energy density of the cell but also introduce additional over-potentials for Mg alloying at the anode. Notably, all binaries considered should be stable vs. Mg metal, except for $MgP_4$, $MgB_4$, and $MgB_7$ (Figure 2), and are candidates for protective coatings at the anode//electrolyte interface. Specifically, Mg-halides including $MgF_2$, $MgCl_2$, $MgBr_2$, which have voltage windows wider than 2.0 V, should be considered as the most promising candidates.

A number of studies have suggested that the $Cl^-$ in magnesium-aluminum-chloride-based electrolytes can protect the Mg-metal anode during Mg deposition via adsorption on the Mg-metal surface[3,4, 29, 56-59] Our results suggest that $MgCl_2$ is stable against the highly reductive environment of Mg-metal, showing a wide stability window ~3.39 V. We speculate that a layer of $MgCl_2$ may form *in situ* as a protective coating, which is further justified by the sparing solubility of this salt in ether-based solvents.[3, 4, 59] Therefore, a careful experimental characterization of the Mg//electrolyte interface will shed light on the role of the speciation of Cl in the form of $MgCl_2$ or as a free ion.

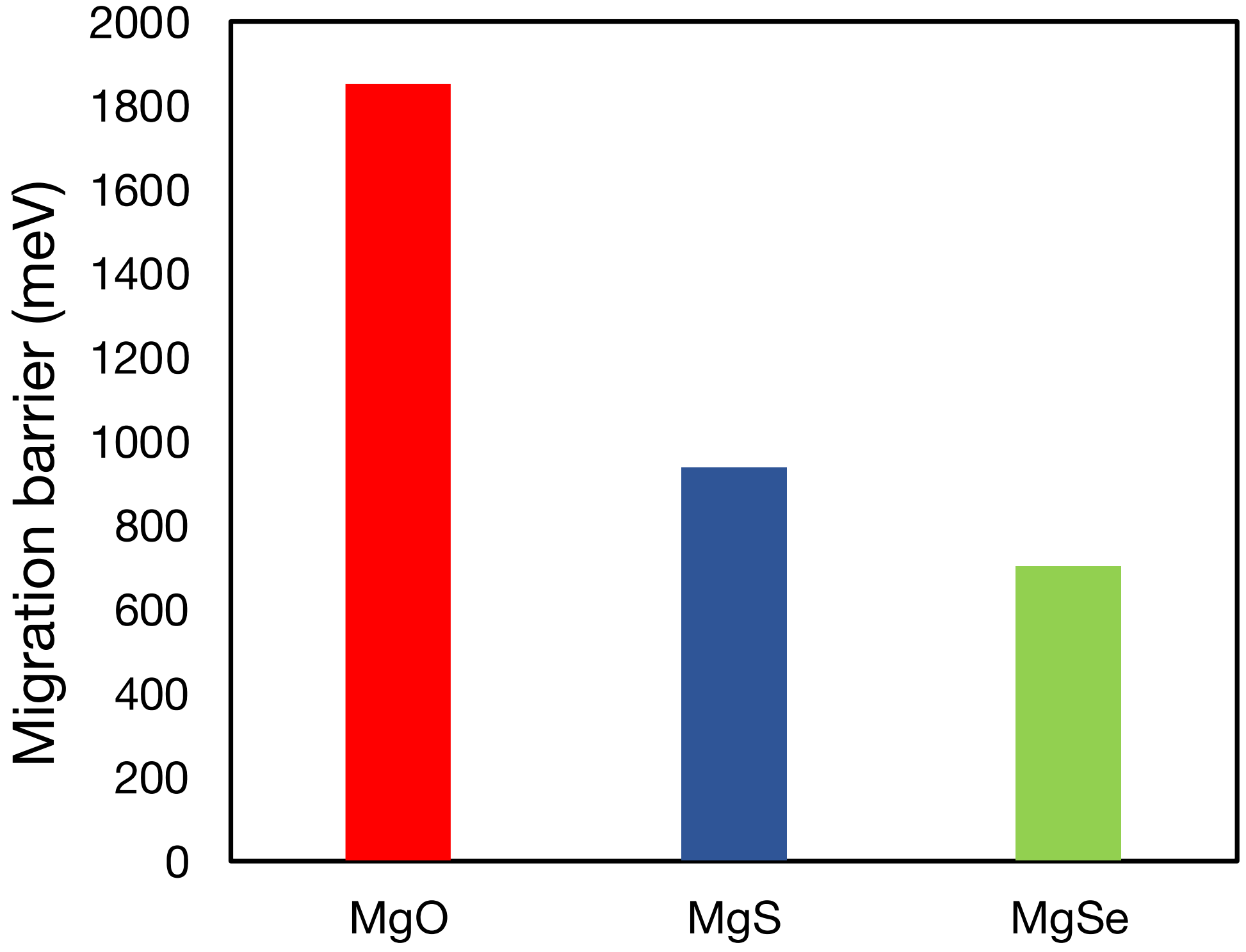

**Figure 5**: Plot of migration barriers of MgO (red), MgS (blue), and MgSe (green) as calculated in Canepa et al.[27] The high migration barriers of MgO, predicted to be stable vs. Mg metal and have a reasonable oxidation limit (3.08 V vs. Mg metal), demonstrate the necessity of $Mg^{2+}$ diffusivity data in determining the viability of potential coating and electrolyte materials.

For all of the suggested anode or cathode coating materials, a thorough evaluation of $Mg^{2+}$ mobility is required to verify their viability as actual coating materials. Mobility evaluations are especially necessary to demonstrate proof-of-concept oxidative coatings that can enable high voltage cathodes (such as $Mg_xCr_2O_4$,[20] $Mg_xMn_2O_4$,[40] and $Mg_xV_2O_5$[51]) in conjunction with current liquid electrolytes and Mg-metal. Note that the $Mg^{2+}$ migration barrier has been calculated for a number of Mg-binaries in a prior study,[27] including MgO (~1800 meV), MgS (~900 meV), and MgSe (~700 meV) of Figure 5, and a few ternaries, such as $Mg(ScSe_2)_2$ (~375 meV), $Mg(InS_2)_2$ (~488 meV) and $Mg(ScS_2)_2$ (~415 meV), while more work is in progress for other candidates listed in this work. The poor bulk Mg mobility causes MgO and MgS to be inactive passivating materials that limit any Mg transference, despite their wide stability ranges (0-3.1 V for MgO and 0-1.6 V for MgS). Similarly, poor Mg mobility in bulk $Mg_{0.5}Ti_2(PO_4)_3$ (> 1 eV[1]) will hinder its use as a protective oxidative coating. Nevertheless, our study identifies a tractable list of possible coating and electrolyte candidates in which $Mg^{2+}$ mobility must be estimated, based on their calculated electrochemical stabilities.

## 5. Conclusion

In this work, we evaluate, using density functional theory calculations, the electrochemical stability windows for non-redox-active Mg binary, ternary and selected quaternary compounds in order to identify potential coating materials for Mg batteries. From the Mg binaries considered, we identify Mg-halides, specifically $MgCl_2$ and $MgBr_2$, as potential anode coating materials based on their reductive stability (at 0 V vs. $Mg/Mg^{2+}$). We also suggest $Mg(BH_4)_2$, $MgAl_2O_4$ and $Mg_2SiO_4$, as possible ternary anode coating materials, given their reductive stability below 0.5 V, with $MgAl_2O_4$ and $Mg_2SiO_4$ exhibiting a voltage window that is >2.0 V wide. Additionally, we expect $MgF_2$, $Mg(PO_3)_2$ and $MgP_4O_{11}$ to be promising candidates for protecting high-voltage cathodes against typical Mg electrolytes. While careful evaluation of Mg mobility in candidate materials is essential, this work identifies specific chemistries as well as general guidelines on compound stabilities that will be useful to design practical coating materials in Mg batteries.

**Acknowledgment**

P.C. acknowledges support from the Singapore Ministry of Education Academic Fund Tier 1 (R-284-000-186-133). This material is based upon work supported by the National Science Foundation Graduate Research Fellowship under Grant No. DGE 1106400. The work is also supported by the Joint Center for Energy Storage Research (JCESR), an Energy Innovation Hub funded by the U.S. Department of Energy, Office of Science and Basic Energy Sciences, through Subcontract 3F-31144. The authors declare no competing financial interests. Any opinion, findings, and conclusions or recommendations expressed in this material are those of the authors and do not necessarily reflect the views of the National Science Foundation.

## References

1. Canepa, P.; Sai Gautam, G.; Hannah, D. C.; Malik, R.; Liu, M.; Gallagher, K. G.; Persson, K. A.; Ceder, G., Odyssey of multivalent cathode materials: open questions and future challenges. *Chemical reviews* **2017,** 117, (5), 4287-4341.
2. Yoo, H. D.; Shterenberg, I.; Gofer, Y.; Gershinsky, G.; Pour, N.; Aurbach, D., Mg rechargeable batteries: an on-going challenge. *Energy & Environmental Science* **2013,** 6, (8), 2265-2279.
3. Doe, R. E.; Han, R.; Hwang, J.; Gmitter, A. J.; Shterenberg, I.; Yoo, H. D.; Pour, N.; Aurbach, D., Novel, electrolyte solutions comprising fully inorganic salts with high anodic stability for rechargeable magnesium batteries. *Chemical Communications* **2014,** 50, (2), 243-245.
4. Canepa, P.; Jayaraman, S.; Cheng, L.; Rajput, N. N.; Richards, W. D.; Gautam, G. S.; Curtiss, L. A.; Persson, K. A.; Ceder, G., Elucidating the structure of the magnesium aluminum chloride complex electrolyte for magnesium-ion batteries. *Energy & Environmental Science* **2015,** 8, (12), 3718-3730.
5. Muldoon, J.; Bucur, C. B.; Gregory, T., Quest for nonaqueous multivalent secondary batteries: magnesium and beyond. *Chemical reviews* **2014,** 114, (23), 11683-11720.
6. Muldoon, J.; Bucur, C. B.; Oliver, A. G.; Sugimoto, T.; Matsui, M.; Kim, H. S.; Allred, G. D.; Zajicek, J.; Kotani, Y., Electrolyte roadblocks to a magnesium rechargeable battery. *Energy & Environmental Science* **2012,** 5, (3), 5941-5950.
7. Hahn, N. T.; Seguin, T. J.; Lau, K.-C.; Liao, C.; Ingram, B. J.; Persson, K. A.; Zavadil, K. R., Enhanced Stability of the Carba-closo-dodecaborate Anion for High-Voltage Battery Electrolytes through Rational Design. *Journal of the American Chemical Society* **2018,** 140, (35), 11076-11084.
8. Aurbach, D.; Lu, Z.; Schechter, A.; Gofer, Y.; Gizbar, H.; Turgeman, R.; Cohen, Y.; Moshkovich, M.; Levi, E., Prototype systems for rechargeable magnesium batteries. *Nature* **2000,** 407, (6805), 724.
9. Cohen, Y. S.; Cohen, Y.; Aurbach, D., Micromorphological studies of lithium electrodes in alkyl carbonate solutions using in situ atomic force microscopy. *The Journal of Physical Chemistry B* **2000,** 104, (51), 12282-12291.
10. Pour, N.; Gofer, Y.; Major, D. T.; Aurbach, D., Structural analysis of electrolyte solutions for rechargeable Mg batteries by stereoscopic means and DFT calculations. *Journal of the American Chemical Society* **2011,** 133, (16), 6270-6278.
11. Mohtadi, R.; Matsui, M.; Arthur, T. S.; Hwang, S. J., Magnesium borohydride: from hydrogen storage to magnesium battery. *Angewandte Chemie International Edition* **2012,** 51, (39), 9780-9783.
12. Carter, T. J.; Mohtadi, R.; Arthur, T. S.; Mizuno, F.; Zhang, R.; Shirai, S.; Kampf, J. W., Boron Clusters as Highly Stable Magnesium‐Battery Electrolytes. *Angewandte Chemie International Edition* **2014,** 53, (12), 3173-3177.
13. Tutusaus, O.; Mohtadi, R.; Arthur, T. S.; Mizuno, F.; Nelson, E. G.; Sevryugina, Y. V., An Efficient Halogen‐Free Electrolyte for Use in Rechargeable Magnesium Batteries. *Angewandte Chemie* **2015,** 127, (27), 8011-8015.

14. Tutusaus, O.; Mohtadi, R.; Singh, N.; Arthur, T. S.; Mizuno, F., Study of electrochemical phenomena observed at the Mg metal/electrolyte interface. *ACS Energy Letters* **2016,** 2, (1), 224-229.
15. Sun, X.; Bonnick, P.; Duffort, V.; Liu, M.; Rong, Z.; Persson, K. A.; Ceder, G.; Nazar, L. F., A high capacity thiospinel cathode for Mg batteries. *Energy Environ. Sci.* **2016,** 9, (7), 2273-2277.
16. Lipson, A. L.; Han, S.-D.; Pan, B.; See, K. A.; Gewirth, A. A.; Liao, C.; Vaughey, J. T.; Ingram, B. J., Practical Stability Limits of Magnesium Electrolytes. *Journal of The Electrochemical Society* **2016,** 163, (10), A2253-A2257.
17. Marom, R.; Amalraj, S. F.; Leifer, N.; Jacob, D.; Aurbach, D., A review of advanced and practical lithium battery materials. *Journal of Materials Chemistry* **2011,** 21, (27), 9938-9954.
18. Goodenough, J. B.; Kim, Y., Challenges for rechargeable Li batteries. *Chemistry of materials* **2009,** 22, (3), 587-603.
19. Lu, Z.; Schechter, A.; Moshkovich, M.; Aurbach, D., On the electrochemical behavior of magnesium electrodes in polar aprotic electrolyte solutions. *Journal of Electroanalytical Chemistry* **1999,** 466, (2), 203-217.
20. Chen, T.; Sai Gautam, G.; Huang, W.; Ceder, G., First-principles study of the voltage profile and mobility of Mg intercalation in a chromium oxide spinel. *Chemistry of Materials* **2017,** 30, (1), 153-162.
21. Liu, M.; Rong, Z.; Malik, R.; Canepa, P.; Jain, A.; Ceder, G.; Persson, K. A., Spinel compounds as multivalent battery cathodes: a systematic evaluation based on *ab initio* calculations. *Energy Environ. Sci.* **2015,** 8, (3), 964-974.
22. Rosenberg, M.; Nicolau, P., Electrical Properties and Cation Migration in $MgMn_2O_4$. **1964,** 101, 101-110.
23. Ling, C.; Zhang, R., Manganese dioxide as rechargeable magnesium battery cathode. *Frontiers in Energy Research* **2017,** 5, 30.
24. Ling, C.; Zhang, R.; Arthur, T. S.; Mizuno, F., How general is the conversion reaction in Mg battery cathode: a case study of the magnesiation of α-MnO2. *Chemistry of Materials* **2015,** 27, (16), 5799-5807.
25. Hannah, D. C.; Sai Gautam, G.; Canepa, P.; Ceder, G., On the Balance of Intercalation and Conversion Reactions in Battery Cathodes. *Advanced Energy Materials* **2018**, 1800379.
26. Gofer, Y.; Turgeman, R.; Cohen, H.; Aurbach, D., XPS investigation of surface chemistry of magnesium electrodes in contact with organic solutions of organochloroaluminate complex salts. *Langmuir* **2003,** 19, (6), 2344-2348.
27. Canepa, P.; Bo, S.-H.; Gautam, G. S.; Key, B.; Richards, W. D.; Shi, T.; Tian, Y.; Wang, Y.; Li, J.; Ceder, G., High magnesium mobility in ternary spinel chalcogenides. *Nature communications* **2017,** 8, (1), 1759.
28. Levi, E.; Gofer, Y.; Aurbach, D., On the way to rechargeable Mg batteries: the challenge of new cathode materials. *Chemistry of Materials* **2009,** 22, (3), 860-868.
29. Canepa, P.; Gautam, G. S.; Malik, R.; Jayaraman, S.; Rong, Z.; Zavadil, K. R.; Persson, K.; Ceder, G., Understanding the initial stages of reversible Mg deposition and stripping in inorganic nonaqueous electrolytes. *Chemistry of Materials* **2015,** 27, (9), 3317-3325.
30. Keyzer, E. N.; Glass, H. F.; Liu, Z.; Bayley, P. M.; Dutton, S. n. E.; Grey, C. P.; Wright, D. S., Mg (PF6) 2-based electrolyte systems: understanding electrolyte–electrode interactions for the development of mg-ion batteries. *Journal of the American Chemical Society* **2016,** 138, (28), 8682-8685.
31. Shao, Y.; Liu, T.; Li, G.; Gu, M.; Nie, Z.; Engelhard, M.; Xiao, J.; Lv, D.; Wang, C.; Zhang, J.-G., Coordination chemistry in magnesium battery electrolytes: how ligands affect their performance. *Scientific reports* **2013,** 3, 3130.

32. Guerfi, A.; Dontigny, M.; Charest, P.; Petitclerc, M.; Lagacé, M.; Vijh, A.; Zaghib, K., Improved electrolytes for Li-ion batteries: Mixtures of ionic liquid and organic electrolyte with enhanced safety and electrochemical performance. *Journal of Power Sources* **2010,** 195, (3), 845-852.
33. Aurbach, D.; Talyosef, Y.; Markovsky, B.; Markevich, E.; Zinigrad, E.; Asraf, L.; Gnanaraj, J. S.; Kim, H.-J., Design of electrolyte solutions for Li and Li-ion batteries: a review. *Electrochimica Acta* **2004,** 50, (2-3), 247-254.
34. Kamaya, N.; Homma, K.; Yamakawa, Y.; Hirayama, M.; Kanno, R.; Yonemura, M.; Kamiyama, T.; Kato, Y.; Hama, S.; Kawamoto, K., A lithium superionic conductor. *Nature materials* **2011,** 10, (9), 682.
35. Masquelier, C., Solid electrolytes: Lithium ions on the fast track. *Nature materials* **2011,** 10, (9), 649.
36. Kato, Y.; Hori, S.; Saito, T.; Suzuki, K.; Hirayama, M.; Mitsui, A.; Yonemura, M.; Iba, H.; Kanno, R., High-power all-solid-state batteries using sulfide superionic conductors. *Nature Energy* **2016,** 1, (4), 16030.
37. Bachman, J. C.; Muy, S.; Grimaud, A.; Chang, H.-H.; Pour, N.; Lux, S. F.; Paschos, O.; Maglia, F.; Lupart, S.; Lamp, P., Inorganic solid-state electrolytes for lithium batteries: mechanisms and properties governing ion conduction. *Chemical reviews* **2015,** 116, (1), 140-162.
38. Verma, P.; Maire, P.; Novák, P., A review of the features and analyses of the solid electrolyte interphase in Li-ion batteries. *Electrochimica Acta* **2010,** 55, (22), 6332-6341.
39. Richards, W. D.; Miara, L. J.; Wang, Y.; Kim, J. C.; Ceder, G., Interface stability in solid-state batteries. *Chemistry of Materials* **2015,** 28, (1), 266-273.
40. Sai Gautam, G.; Canepa, P.; Urban, A.; Bo, S.-H.; Ceder, G., Influence of inversion on Mg mobility and electrochemistry in spinels. *Chemistry of Materials* **2017,** 29, (18), 7918-7930.
41. Canepa, P.; Sai Gautam, G.; Broberg, D.; Bo, S.-H.; Ceder, G., Role of point defects in spinel Mg chalcogenide conductors. *Chemistry of Materials* **2017**.
42. Jain, A.; Hautier, G.; Moore, C. J.; Ong, S. P.; Fischer, C. C.; Mueller, T.; Persson, K. A.; Ceder, G., A high-throughput infrastructure for density functional theory calculations. *Computational Materials Science* **2011,** 50, (8), 2295-2310.
43. Ong, S. P.; Richards, W. D.; Jain, A.; Hautier, G.; Kocher, M.; Cholia, S.; Gunter, D.; Chevrier, V. L.; Persson, K. A.; Ceder, G., Python Materials Genomics (pymatgen): A robust, open-source python library for materials analysis. *Computational Materials Science* **2013,** 68, 314-319.
44. Jain, A.; Ong, S. P.; Hautier, G.; Chen, W.; Richards, W. D.; Dacek, S.; Cholia, S.; Gunter, D.; Skinner, D.; Ceder, G., Commentary: The Materials Project: A materials genome approach to accelerating materials innovation. *Apl Materials* **2013,** 1, (1), 011002.
45. Hohenberg, P., Kohn, W., Inhomogeneous electron gas. *Physical Review B* **1973,** 7, (5), 1912-1919.
46. Kohn, W.; Sham, L. J., Self-consistent equations including exchange and correlation effects. *Physical Review* **1965,** 140, (4A).
47. Bergerhoff, G.; Brown, I., Inorganic crystal structure database. **1987**.
48. Hart, G. L.; Forcade, R. W., Algorithm for generating derivative structures. *Physical Review B* **2008,** 77, (22), 224115.
49. Hart, G. L.; Forcade, R. W., Generating derivative structures from multilattices: Algorithm and application to hcp alloys. *Physical Review B* **2009,** 80, (1), 014120.
50. Hart, G. L.; Nelson, L. J.; Forcade, R. W., Generating derivative structures at a fixed concentration. *Computational Materials Science* **2012,** 59, 101-107.

51. Sai Gautam, G.; Canepa, P.; Abdellahi, A.; Urban, A.; Malik, R.; Ceder, G., The intercalation phase diagram of Mg in $V_2O_5$ from first-principles. *Chemistry of Materials* **2015,** 27, (10), 3733-3742.
52. Pauling, L., THE NATURE OF THE CHEMICAL BOND. IV. THE ENERGY OF SINGLE BONDS AND THE RELATIVE ELECTRONEGATIVITY OF ATOMS. *Journal of the American Chemical Society* **1932,** 54, (9), 3570-3582.
53. Arthur, T. S.; Glans, P.-A.; Singh, N.; Tutusaus, O.; Nie, K.; Liu, Y.-S.; Mizuno, F.; Guo, J.; Alsem, D. H.; Salmon, N. J., Interfacial Insight from Operando XAS/TEM for Magnesium Metal Deposition with Borohydride Electrolytes. *Chemistry of Materials* **2017,** 29, (17), 7183-7188.
54. Ma, C.; Cheng, Y.; Yin, K.; Luo, J.; Sharafi, A.; Sakamoto, J.; Li, J.; More, K. L.; Dudney, N. J.; Chi, M., Interfacial stability of Li metal–solid electrolyte elucidated via in situ electron microscopy. *Nano letters* **2016,** 16, (11), 7030-7036.
55. Arthur, T. S.; Singh, N.; Matsui, M., Electrodeposited Bi, Sb and Bi1-xSbx alloys as anodes for Mg-ion batteries. *Electrochemistry Communications* **2012,** 16, (1), 103-106.
56. Aurbach, D.; Gizbar, H.; Schechter, A.; Chusid, O.; Gottlieb, H. E.; Gofer, Y.; Goldberg, I., Electrolyte solutions for rechargeable magnesium batteries based on organomagnesium chloroaluminate complexes. *Journal of The Electrochemical Society* **2002,** 149, (2), A115-A121.
57. See, K. A.; Chapman, K. W.; Zhu, L.; Wiaderek, K. M.; Borkiewicz, O. J.; Barile, C. J.; Chupas, P. J.; Gewirth, A. A., The interplay of Al and Mg speciation in advanced Mg battery electrolyte solutions. *Journal of the American Chemical Society* **2015,** 138, (1), 328-337.
58. See, K. A.; Liu, Y.-M.; Ha, Y.; Barile, C. J.; Gewirth, A. A., Effect of Concentration on the Electrochemistry and Speciation of the Magnesium Aluminum Chloride Complex Electrolyte Solution. *ACS applied materials & interfaces* **2017,** 9, (41), 35729-35739.
59. Salama, M.; Shterenberg, I.; JW Shimon, L.; Keinan-Adamsky, K.; Afri, M.; Gofer, Y.; Aurbach, D., Structural Analysis of Magnesium Chloride Complexes in Dimethoxyethane Solutions in the Context of Mg Batteries Research. *The Journal of Physical Chemistry C* **2017,** 121, (45), 24909-24918.